%% file: BC.tex
\documentclass[pdf]{PoS}
\usepackage{paralist}

\def\Url#1{\href{#1}{\texttt{[#1]}}}

\newcommand{\alphas}{\ensuremath{\alpha_{\mathrm{s}}}}
\newcommand{\smgg}{\ensuremath{\mathrm{SU(3)_c} \otimes \mathrm{SU(2)_L} \otimes \mathrm{U(1)}_Y}}
\newcommand{\cgg}{\ensuremath{\mathrm{SU(3)_c}}}

\newcommand{\ewgg}{\ensuremath{\mathrm{SU(2)_L} \otimes \mathrm{U(1)}_Y}}

\newcommand{\gev}{\ensuremath{\hbox{ GeV}}}
\newcommand{\tev}{\ensuremath{\hbox{ TeV}}}
\newcommand{\fb}{\ensuremath{\hbox{ fb}}}
\newcommand{\cm}{\ensuremath{\hbox{ cm}}}
\newcommand{\etal}{\textit{et al.}}

\newcommand{\mev}{\ensuremath{\hbox{ MeV}}}

\renewenvironment{itemize}[1]{\begin{compactitem}#1}{\end{compactitem}}
\newenvironment{packed_enum}{
\begin{enumerate}
  \setlength{\itemsep}{0pt}
  \setlength{\parskip}{0pt}
  \setlength{\parsep}{0pt}
  \setlength{\partopsep}{0pt}
}{\end{enumerate}}

\input{BCjournals}
\graphicspath{{ConfXfigs/}}

\title{Beyond Confinement}

\ShortTitle{Beyond Confinement}

\author{\speaker{Chris Quigg}\\
        Fermi National Accelerator Laboratory\thanks{Operated by Fermi Research Alliance, LLC under Contract No.~DE-AC02-07CH11359 with the United States Department of Energy.}, P.O. Box 500, Batavia, Illinois 60510 USA\\
        E-mail: \email{quigg@fnal.gov}}

\abstract{A digest of my closing remarks at ConfX.\hfill {\fbox{\textsf{FERMILAB-CONF-13/008-T}}}}

\FullConference{Xth Quark Confinement and the Hadron Spectrum,\\
		October 8-12, 2012\\
		TUM Campus Garching, Munich, Germany}

\begin{document}

\section{Is QCD a Confining Theory?}
Eighteen years after the inaugural meeting in this series, it is worth noting that color confinement has not yet been proved to the standards of axiomatic field theory. In their introduction to the Clay Mathematical Institute's Millennium Prize Problem on Yang--Mills theory, Arthur Jaffe and Edward Witten note~\cite{claymation} that for quantum chromodynamics to describe the strong interaction successfully, in contrast to the classical non-Abelian gauge theory:
\begin{packed_enum}
\item It must have a mass gap.
\item It must display confinement, so that only color singlets are ``physical particle states.''
\item It must manifest chiral symmetry breaking.
\end{packed_enum}
Accordingly, they set the (million-dollar) challenge,
\begin{quote}
\textsc{Yang--Mills existence and Mass Gap.} \emph{Prove that for any compact simple gauge group $G$, a non-trivial Yang--Mills theory exists on $R^4$ and has a mass gap $\Delta > 0$. Existence includes establishing axiomatic properties at least as strong as  \ldots }
\end{quote}
Lattice gauge theory and other nonperturbative approaches have given us much circumstantial evidence for---and insight into---the three crucial properties. Certainly it is a very credible working hypothesis to suppose the QCD is a confining theory. It is nevertheless worth acknowledging that, at a foundational level, the case remains open.

\section{Asymptotic Freedom and the Origins of Hadron Mass}
\begin{figure}[bt] 
	\centerline{\includegraphics[width=8.5cm]{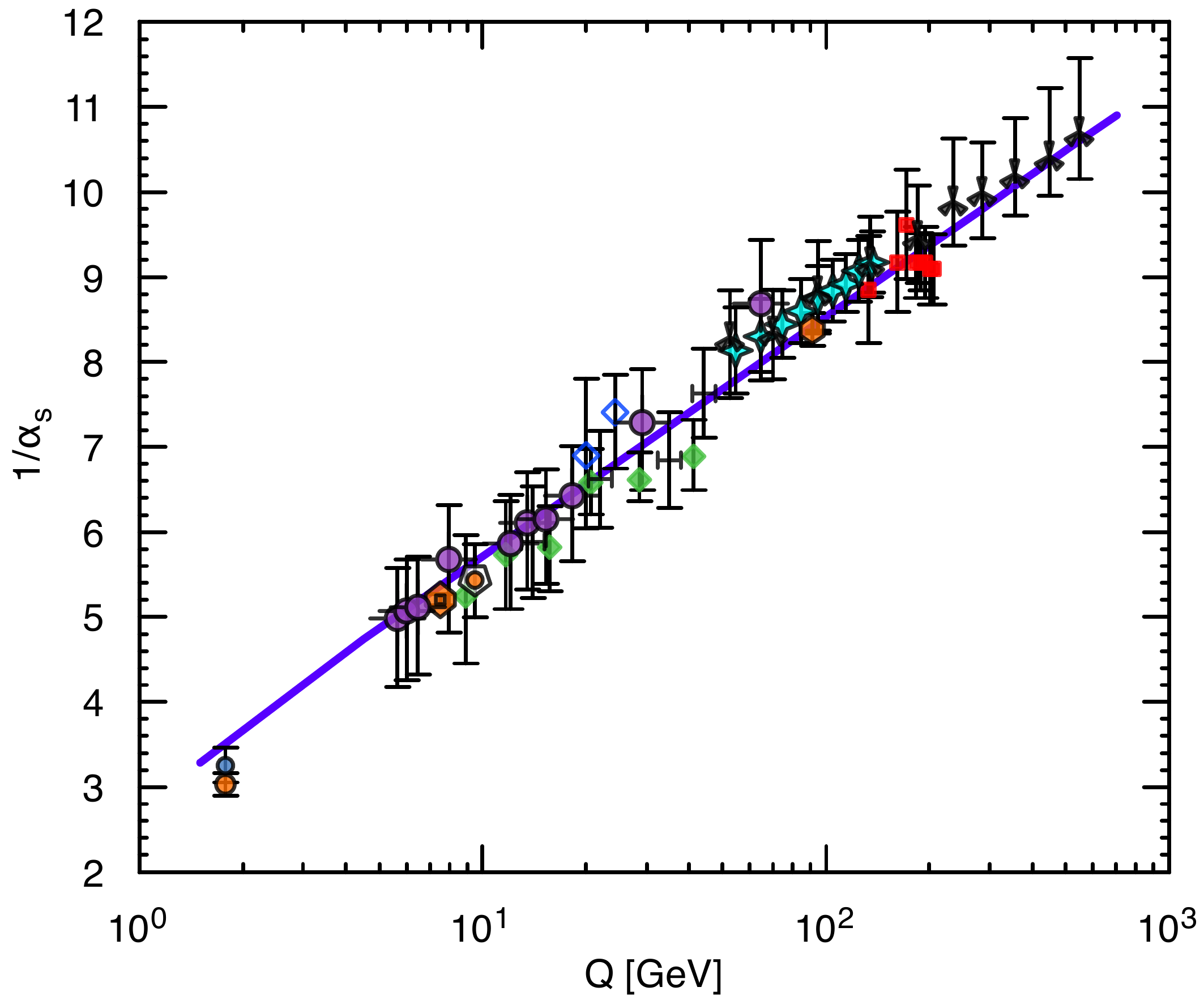}}
	\caption{Measurements of the strong coupling $1/\alphas(Q)$ as a
		function of the energy scale $\ln{Q}$.
        In addition to hadronic $\tau$-decay, quarkonium, $\Upsilon$ 
        decay, and $Z^0$-pole values, 
        we display black crosses: $e^+e^-$ collisions;
        red squares: $e^+e^-$ collisions; 
        green diamonds: $e^\pm p$ collisions;
        barred purple circles: $e^\pm p$ collisions; 
        cyan crosses: $\bar{p}p$ collisions;
		average value of $\alphas(M_Z)$, in 4-loop approximation and
		using 3-loop threshold matching at the heavy-quark pole masses
		$m_c = 1.5\gev$ and $m_b = 4.7\gev$. From Ref.~\cite{Kronfeld:2010bx}{, extended to high scales by the determinations of $\alphas$ from ATLAS jet data~\cite{Malaescu:2012ts}.}
}
	\label{fig:kqasf}
\end{figure}

Asymptotic freedom, the tendency within QCD of the strong coupling $\alphas(Q)$ to decrease with increasing scale $Q$, is an experimental fact. At lowest nontrivial order, we expect $1/\alphas(Q)$ to increase linearly with $\ln{Q}$. That this is true to excellent approximation is shown in the compilation plotted in Figure~\ref{fig:kqasf}.

Confinement, expressed through the quantitative instrument of lattice QCD, explains nearly all of the mass of the light hadrons in our world~\cite{Kronfeld:2012ym}, and so explains nearly all the visible mass in the universe. Indeed, the nucleon mass is a prime exemplar of Einstein's original formulation as $m = E_0/c^2$, where $E_0$ is the rest energy of a body and $c$ is the speed of light. The up and down quarks contribute only a few percent of the  isoscalar nucleon mass (939\mev), because~\cite{pdg12}
\begin{equation}
3\,\frac{m_u + m_d}{2} \approx 9.6\hbox{ to }13.2\mev.
\label{eq:lqms}
\end{equation}
Chiral perturbation theory tells us that in the limit of vanishing quark masses the nucleon mass would decrease to $M_N \approx 870\mev$~\cite{chiptMN}. A small real-world contribution
from the strange-quark sea would be absent.
Of course, we need the light-quark masses to explain $M_p < M_n$, a defining aspect of the real world.

From the perspective of unified theories, quark masses do matter in other ways because their values are encoded in the low-energy values of the \smgg\ couplings. For example, it is easy to see how the top-quark mass influences the low-energy value of \alphas~\cite{topology}.
In unified theories of the strong, weak, and electromagnetic 
in\-ter\-ac\-tions, all the  coupling ``constants'' take on a 
common value, $\alpha_{\mathrm{U}}$, at some high energy, $M_{\mathrm{U}}$.  If we adopt the 
point of view that
$\alpha_{\mathrm{U}}$ is fixed at the unification 
scale, then the mass of the top quark is encoded in the value of 
the strong coupling $\alphas$ that we 
experience at low energies.  Assuming three generations of quarks and 
leptons, we  evolve $\alphas$ downwards in energy from the 
unification scale.
The leading-logarithmic behavior is given by
\begin{equation}
1/\alphas(Q) = 1/\alpha_{\mathrm{U}} + \frac{21}{6\pi}\ln(Q/M_{\mathrm{U}})\;\; ,
\end{equation} for $M_{\mathrm{U}} > Q > 2 m_t$.  The positive coefficient 
$+21/6\pi$ means that the strong coupling constant 
$\alphas$ is smaller at high energies than at low energies. [The entry at a scale $\widetilde{Q}$ of a full complement of superpartners would change the slope from $21/6\pi = 7/2\pi$ to $3/2\pi$.]

\begin{figure}[tb]
\centerline{\includegraphics[width=0.6\textwidth]{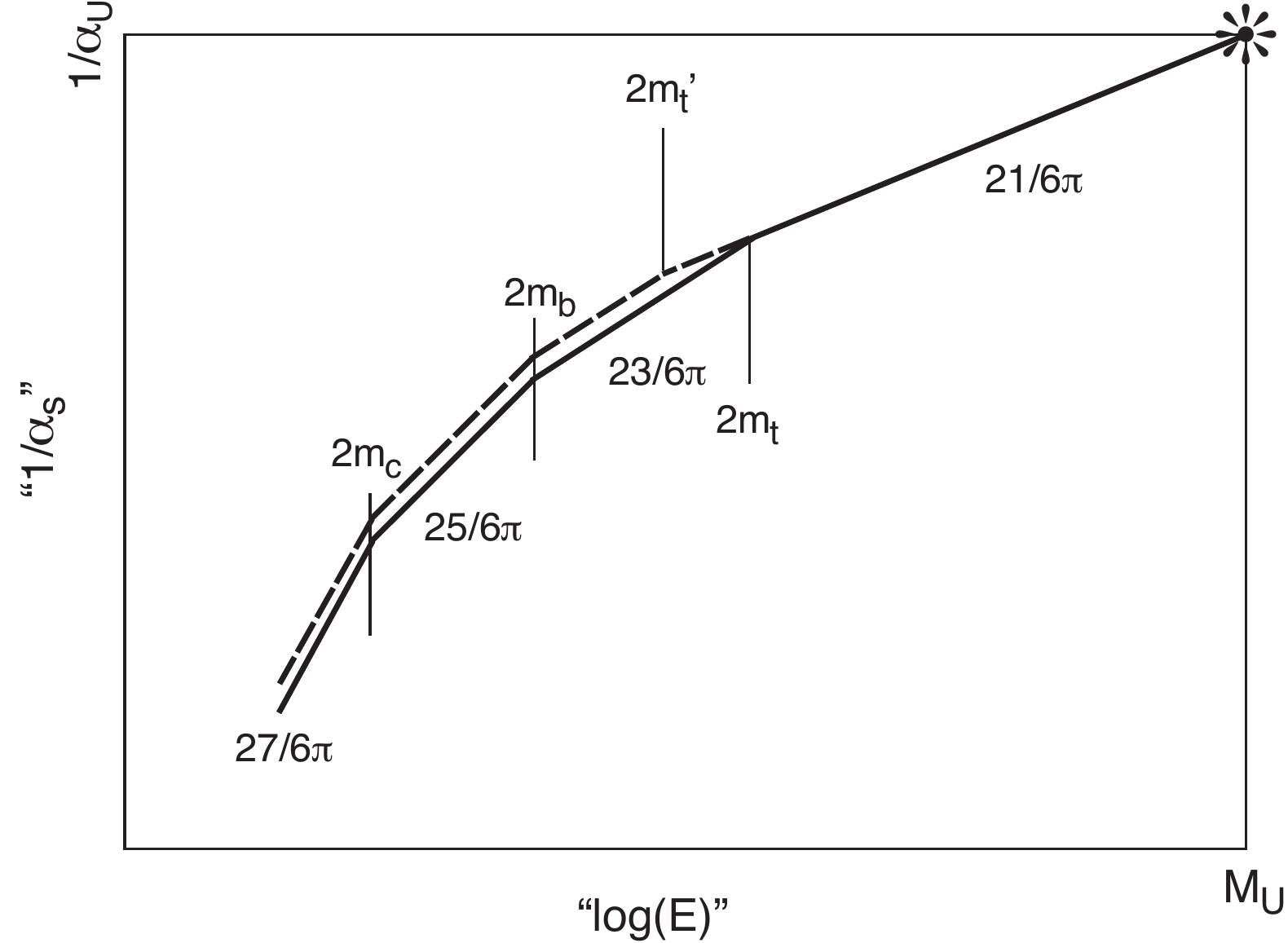}}
\caption{Two evolutions of the strong coupling, \alphas. A smaller value of the top-quark mass leads to a small value of \alphas.}
\label{fig:runningas}
\end{figure}

In the interval between $2m_t$ and $2m_b$, the 
slope $(33-2n_{\!f})/6\pi$ (where $n_{\!f}$ is the number of active quark 
flavors) steepens to $23/6\pi$, and then increases by 
another $2/6\pi$ at every quark threshold.  This behavior is 
shown by the solid line in Figure~\ref{fig:runningas}.
The dotted line in Figure \ref{fig:runningas} shows how the evolution of 
$1/\alphas$ changes if the top-quark mass is reduced.  A smaller top 
mass means a larger low-energy value of $1/\alphas$, so a smaller 
low-energy value of $\alphas$.  

Neglecting the tiny  light-quark masses, the scale parameter $\Lambda_{\hbox{\footnotesize QCD}}$ is the 
only dimensionful parameter in QCD.  It determines the scale of the 
confinement energy that is the dominant contribution to the nucleon mass. 
To a good first approximation, 
\begin{equation}
	M_N \approx C \Lambda_{\hbox{{\footnotesize QCD}}},
	\label{lattice}
\end{equation}
where the constant of proportionality $C$ is calculable using 
techniques of lattice field theory.
 
To discover the dependence of $\Lambda_{\hbox{{\footnotesize QCD}}}$ upon the top-quark 
mass, we calculate $\alphas(2m_t)$ 
evolving up from low energies and down from the unification scale, and match:
\begin{equation}
1/\alpha_U  +  {\displaystyle \frac{21}{6\pi}}\ln(2m_t/M_U) =  
 1/\alphas(2m_c) - {\displaystyle \frac{25}{6\pi}}\ln(m_c/m_b) 
 -{\displaystyle \frac{23}{6\pi}}\ln(m_b/m_t)   .
 \end{equation} Identifying
 \begin{equation}
 1/\alphas(2m_c) \equiv {\displaystyle\frac{ 
 27}{6\pi}}\ln(2m_c/\Lambda_{\hbox{{\footnotesize QCD}}})\;  ,
\end{equation}  we find that 
\begin{equation}
	\Lambda_{\hbox{{\footnotesize QCD}}}=e^{\displaystyle -6\pi/27\alpha_U} 
	\left(\frac{M_U}{1 \gev}\right)^{\!21/27} 
	\left(\frac{2m_t\cdot 2m_b\cdot 2m_c}{1\gev^{3}}\right)^{\!2/27}\gev \;\; .
	\label{blank}
\end{equation}  
We conclude that, in a simple unified theory,
\begin{equation}
	\frac{M_N}{1\gev} \propto 
	\left(\frac{m_t}{1\gev}\right)^{2/27} \;\; .
	\label{amazing}
\end{equation}

I invite you to consider what happens to the nucleon mass---in the unified-theory framework---if {all} the quark masses are taken to be very small and, if you choose, to look beyond the leading-logarithmic evolution of \alphas. Naturally, if the quark masses all vanish, \emph{all} of the isoscalar nucleon mass arises from confinement, though it will be roughly half what it is in the real world.

\section{Is QCD Complete?}
Frank Wilczek has proclaimed QCD ``our most perfect physical theory,'' in part because it lacks structural flaws that would cause it to break down at high energies~\cite{Wilczek:1999id}. Just because the theory could be complete all the way up to the Planck scale doesn't mean that it will prove to be the final word. So while it makes sense to treat QCD with  respect and to rely on it to calculate backgrounds and signals, let us remain open to surprises. Cracks in (our current understanding of) QCD could appear as breakdowns in factorization, compromising our ability to make reliable perturbative predictions, the observation of free quarks or unconfined color, novel forms of colored matter, a larger symmetry containing \cgg, or even quark compositeness.

In my view, it is more likely that we will encounter new phenomena \emph{within} QCD that do not shake the foundations, but give us new ``applied science'' questions to contemplate, and help us understand the full richness of the theory. At LHC energies, will ``soft'' multiparticle production exhibit new mechanisms beyond diffraction and multiperipheral-style short-range order? Will the expected high density of few-GeV partons lead to thermalization, revealed especially in high-multiplicity events~\cite{Shuryak}, or to events containing many minijets? Without any specific basis, I suspect that a few percent of ``minimum-bias'' or ``zero-bias'' events will display new event structures that become increasingly prominent with increasing collision energy and multiplicity. Bjorken argues that we might be able to identify classes of collisions that reveal distinct body plans for the proton, including quark--diquark  or diquark--diquark encounters. The apparently novel long-range correlations in rapidity reported by the CMS collaboration in $pp$ and $p$-Pb collisions~\cite{Khachatryan:2010gv,CMS:2012qk} may be harbingers of new collective dynamics.

\section{The Strong \textsf{CP} Problem}
In contrast to the electroweak theory, which is menaced by vacuum-stability and triviality concerns unless the Higgs-boson mass lies in the interval $134\gev \lesssim M_H \lesssim 177\gev$, quantum chromodynamics could be internally consistent up to very high scales, such as the unification scale or the Planck scale. There is, however, one pea ($\mathsf{CP}$) under our mattress~\cite{HCApea}: the threat to  $\mathsf{CP}$-invariance posed by the $\theta$-vacuum of QCD~\cite{hooft}. The phase of the quark mass matrix and the coefficient of the $G\,^{*\!}G$ term in the QCD Lagrangian---two quantities with distinctly separate origins---combine to cause effects that violate $\mathsf{CP}$ invariance. 

To respect the upper limit on the neutron's electric dipole moment~\cite{ndnlim},
\begin{equation}
|d_n| \lesssim 0.29 \times 10^{-25}~e\cm \hbox{ at 90\% CL,}
\label{c8:ndn}
\end{equation}
the effective parameter $\theta$ must be extraordinarily small~\cite{thetlim},
\begin{equation}
|\theta| \lesssim 10^{-10} .
\label{c8:thetlim}
\end{equation}
The mystery of the exquisite smallness of $\theta$ is the \textit{strong} $\mathsf{CP}$ \textit{problem.} The most promising strategy consists in adding a second Higgs doublet and an additional $\mathrm{U(1)}$ symmetry to the standard model Lagrangian~\cite{pecceiq}. The new $\mathrm{U(1)}$ symmetry is spontaneously broken in the course of electroweak symmetry breaking, and minimizing the Higgs potentials enforces $\theta=0$. The breaking of the $\mathrm{U(1)}$ symmetry implies the existence of a new pseudo-Nambu--Goldstone boson, called the axion, with several implications for particle physics and cosmology~\cite{axion,Kim:2008hd}. No signal for an axion has yet been found; many imaginative searches are ongoing~\cite{patras2012}. Axions could also account for some or all of the dark matter of the universe~\cite{Duffy:2009ig}.

\section{Direct Dark Matter Searches}
What instruments we have agree that the universe is composed of
23.6\% dark matter, 4.6\% ordinary (baryonic) matter, and 71.8\% dark energy~\cite{Bennett:2012fp}. For the moment at least, exploring the nature of dark energy lies in the realm of astronomical observations. The search for dark matter, on the other hand, is everyday business for particle physicists. On its face, the search for dark matter entails physics beyond the standard model---but understanding nucleon structure may be a key to interpreting results. In an interesting class of supersymmetric scenarios, the scattering of a weakly interacting massive particle from nucleons is mediated by Higgs-boson exchange~\cite{Baltz:2006fm}. Quantifying the Higgs-boson interaction with the nucleon calls for a good understanding of the heavy-flavor content of the nucleon, itself a prime concern of hadronic physics over many years~\cite{Thomas:2012tg}. Over the past decade, increasingly plausible lattice-QCD evaluations of the heavy-quark content of the nucleon have complemented the experimental program~\cite{Freeman:2012ry}. These efforts acquire added urgency as WIMP searches enter the domain in which supersymmetric signals are anticipated~\cite{Aprile:2012nq}. Interpretation of search results will benefit from sharper estimates of the WIMP--nucleon cross section.

The direct searches that are sensitive to collisions of target nucleons with WIMPS as Earth moves through the bath of dark-matter particles are complemented by generic collider searches for missing transverse energy signatures~\cite{paddy,lhcdm}. Here, too, QCD is essential to the computation of both backgrounds and putative signals.

\section{The Electroweak Theory and the Higgs Boson}
Our standard electroweak theory is based on a three-generation, $\mathrm{V - A}$ description of the charged-current interactions, in which the (Cabibbo--Kobayashi--Maskawa) quark-mixing matrix describes \textsf{CP} violation. The Glashow--Iliopoulos--Maiani mechanism~\cite{Glashow:1970gm} controls flavor-changing neutral currents. Although the \ewgg\ gauge symmetry is hidden, it has been validated in measurements of the reaction $e^+e^- \to W^+W^-$ at LEP~\cite{LEPWW}. The electroweak theory has been tested as a quantum field theory at the per-mille level~\cite{ALEPH:2005ab}. The nature of the agent that hides electroweak symmetry has been a leading mystery for decades.
I cite what have been four of the leading possibilities:
\begin{itemize}
\item[ ] A force of a new character, based on interactions of a fundamental scalar (the standard-mdoel possibility);
\item[ ] A new gauge force, perhaps acting on undiscovered constituents;
\item [ ] A residual force that emerges from strong dynamics among electroweak gauge bosons;
\item[ ] An echo of extra spacetime dimensions.
\end{itemize}

\subsection{(To What Extent) Have We Found the Higgs Boson?}
In July 2012,  ATLAS~\cite{atlashiggs:2012gk} and CMS~\cite{cmshiggs:2012gu}  reported the discovery at $5\sigma$ significance of a neutral, even-integer-spin, particle in the neighborhood of $125\hbox{ - }126\gev$. The product of production cross section times branching fraction for the $\gamma\gamma$, $ZZ^*$, and $WW^*$ channels is consistent, within limited statistics, with expectations for the standard-model Higgs boson. We expect more definitive statements based on the full 2012 data set within the next half year~\cite{science2012}. The CDF and D0 Collaborations combined their searches for a Higgs boson that decays into $b\bar{b}$, produced in association with  $W^\pm$ or $Z^0$,  in up to $9.7\fb^{-1}$ in 1.96-TeV $\bar{p}p$ collisions. They report an excess over background that reaches a significance of $3.1\sigma$ in the mass range between 120 and $135\gev$~\cite{Aaltonen:2012qt}.

Among the urgent questions are these: What are the quantum numbers of the new state? Does it fully account for electroweak symmetry breaking, as reflected in its couplings to $W$ and $Z$? Does it couple to fermions (qualitatively), and does it account quantitatively for the fermion masses. Do the branching fractions into $W^+W^-$, $Z^0Z^0$, and $\gamma\gamma$ match standard-model expectations? Are the production modes as anticipated? Does it decay into any hitherto unknown particles? Is there any sign of new strong dynamics~\cite{Eichten:2012qb}? What are the implications of $M_H \approx 126\gev$?

\subsection{Why Does Finding the Higgs Boson Matter?}
It is worth asking how different the world would have been without a mechanism for spontaneous symmetry breaking at the electroweak scale~\cite{Quigg:2009xr}. Many of the essential features can be distilled from a toy model with a single generation of quarks and leptons. Without a nonzero vacuum expectation value for a Higgs field, the up and down quarks and the electron would be massless. QCD would still combine the color-triplet quarks into color singlets, and the gross features of nucleons---such as masses---would be little changed. (However, the proton might outweigh the neutron. We found the question too close to call.) The $\mathrm{SU(2)_L \otimes SU(2)_R}$ chiral symmetry that the QCD Lagrangian displays for massless quarks is broken to isospin symmetry near the confinement scale by the formation of $\langle\bar{q}q\rangle = \langle \bar{q}_{\mathrm{L}}q_{\mathrm{R}} \rangle + \langle \bar{q}_{\mathrm{R}}q_{\mathrm{L}} \rangle$ condensates. By coupling left-handed and right-handed quarks, the condensates break the electroweak symmetry and produce effective ``constituent-quark'' masses. The weak bosons acquire masses, but they are $2\,500$ times smaller than in the real world. The analogue of the Fermi constant is $\sim 10^7 G_{\mathrm{F}}$. 

If the proton were stable or if compound nuclei were produced and survived until late times in this hypothetical world, we could entertain the possibility of atoms. But the infinitesimal (even vanishing) electron mass means that the Bohr radius of a would-be atom would be macroscopic, if not infinite. It would be impossible to associate a specific electron with a particular nucleus, so valence bonding would have no meaning. As we characterize the agent that breaks electroweak symmetry, we seek to learn why atoms and chemistry and solids and liquids exist---why the everyday world is as we find it. Whoever shows (presumably in the far future) how the \textit{electron} acquires mass will merit a Nobel Prize in Chemistry!

\section{What Sets Fermion Masses?}
According to physics tradition, Feynman kept in the corner of his blackboard the question, ``Why does the muon weigh?'' 
Suppose that our experiments are able to demonstrate that the Higgs-boson candidate observed at the LHC couples to fermions with strength $m_f/v$, where $m_f$ is the fermion mass and $v \approx 246\gev$ is the vacuum expectation value of the Higgs field. Then we can conclude that the fermion mass arises from a Yukawa term in the electroweak Lagrangian. For the muon, the form would be
\begin{displaymath}
\mathcal{L}_{\mathrm{Yukawa}} = -\zeta_\mu\left[\bar{\mu}_{\mathsf{R}}(\phi^\dagger \mu_{\mathsf{L}}) + (\bar{\mu}_{\mathsf{L}}\phi)\mu_{\mathsf{R}}\right],
\label{eq:yuk}
\end{displaymath}
and appropriate generalizations would account for mixing, in the case of quarks. Then we might claim to answer Feynman's question in a restricted but important sense: we will understand how the fundamental fermions acquire mass. A big question remains: what sets the value of the fermion masses---``\emph{What} do the fermions weigh?''

Within the standard model, the Yukawa couplings are simply chosen to reproduce the observed masses; they are not predicted. I therefore regard all fermion mass---starting with the electron mass---as physics beyond the standard model, even if the Higgs mechanism indeed lies behind the existence of quark and charged-lepton masses. Neutrino masses may entail additional new physics.

\section{The Unreasonable Effectiveness of the Standard Model}
Whereas the GIM mechanism~\cite{Glashow:1970gm} inhibits flavor-changing neutral currents  (FCNC) within the standard electroweak theory, extensions to the standard model generically give rise to FCNC. An important issue for Technicolor, for many supersymmetric extensions to the electroweak theory, and for other ``new physics'' models has been to find graceful ways of enforcing limitations on FCNC, to survive experimental constraints. Nevertheless, we have expected the absence of FCNC to be an idealization, and have sought evidence of FCNC well above the standard-model level in characteristic settings. 

Somewhat surprisingly, we have no evidence for nonstandard FCNC anywhere. The most recent experimental progress concerns the decay $B_s \to \mu^+\mu^-$ in which a class of supersymmetric models foresee a considerable amplification of the standard-model rate. The LHC experiment has now made the first observation of this decay~\cite{stone} at a branching fraction, $B(B_s \to \mu^+\mu^-) = (3.2^{+1.5}_{-1.2})\times10^{-9}$. in accord with the standard-model expectation, $(3.54 \pm 0.30)\times10^{-9}$~\cite{Buras:2012ru}. For a commentary on implications for supersymmetry (which continues to hide very effectively), see Ref.~\cite{herbi}.

I take the continued absence of FCNC as a powerful suggestion that we are missing some dynamical principle or symmetry. One framework for dealing with the data is ``minimal flavor violation''~\cite{D'Ambrosio:2002ex}. Proposals to account for the fermion mass spectra and mixing angles must attend to the implications for FCNC. Among the strategies under active study are new implementations of the Froggat--Nielsen mechanism~\cite{Babu:2009fd} and studies of partially composite Higgs bosons~\cite{Kaplan:1983fs}.

 \section*{Acknowledgements and Thanks}
Conversations with Lance Dixon, Andreas Kronfeld, and Robert Shrock have shaped my perspectives on the influence of quark masses on the proton mass. I thank Bogdan Malaescu for providing \alphas\ values from Ref.~\cite{Malaescu:2012ts}. 
 I am grateful to Nora Brambilla, Andrzej Buras, Stephan Paul, and the Cluster of Excellence for Fundamental Physics: \emph{Origin and Structure of the Universe} for warm hospitality in Munich. It is a pleasure to thank all the ConfX organizers and participants for a stimulating and lively meeting.

\end{document}

%% file: BCjournals.tex
\def\pl#1#2#3{{Phys. Lett. }{\bf #1} (19#3) #2}
        
\def\prl#1#2#3{{Phys. Rev. Lett. }{\bf #1} (19#3) #2}

\def\pr#1#2#3{{Phys. Rev. D}{\bf #1} (19#3) #2}

\def\sciam#1#2#3#4{{\it Sci. Am. }{\bf #1}, #2 (\ifcase#3\or January\or February\or March\or April\or May\or
June\or July\or August\or September\or October\or November\or December\fi, 19#4)}

\def\phystoday#1#2#3#4{{\it Phys. Today }{\bf #1}, #2 (\ifcase#3\or January\or 
         February\or March\or April\or May\or June\or July\or August\or 
         September\or October\or November\or December\fi, 19#4)}
        
\relax